\documentclass[onecolumn, usenatbib]{mnras}
\usepackage{times,graphicx,amssymb, amsmath, natbib}
\usepackage[T1]{fontenc}

\usepackage{epsfig}

\newcommand{\beq}{\begin{equation}}
\newcommand{\eeq}{\end{equation}}

\newcommand{\ber}{\begin{eqnarray}}
\newcommand{\eer}{\end{eqnarray}}

\def\apj{{Astroph.\@ J.\ }}
\def\mnras{{Mon.\@ Not.\@ Roy.\@ Ast.\@ Soc.\ }}

\def\aj{{Astron.\@ J.\ }}

\def\prd{{Phys.\@ Rev.\@ D\ }}

\def\beq{\begin{equation}}
\def\eeq{\end{equation}}

\def\ber{\begin{eqnarray}}
\def\eer{\end{eqnarray}}

\begin{document}

\title[Probing quintessence with  HI  intensity mapping]
{Prospects of probing quintessence with  HI 21-cm  intensity mapping survey}

\author[Hussain etal.]
{Azam Hussain $^{1}$\thanks{E-mail:azam@ctp-jamia.res.in}, Shruti Thakur$^{1}$\thanks{shruti@ctp-jamia.res.in}, 
Tapomoy Guha Sarkar $^{2}$ \thanks{E-mail: tapomoy1@gmail.com}, 
Anjan A Sen $^{1}$ \thanks{E-mail:aasen@jmi.ac.in} \\
$^{1}$Centre for Theoretical Physics, Jamia Millia Islamia, New Delhi-110025, India.\\
$^{2}$Department of Physics, Birla Institute of Technology and Science, Pilani, Rajasthan, 333031. India.
}

\maketitle
\date{\today}

\begin{abstract}
We investigate the prospect of constraining scalar field dark energy
models using HI 21-cm intensity mapping surveys.  We consider a wide class
of coupled scalar field dark energy models whose predictions about the background cosmological evolution are different from the $\Lambda$CDM
predictions by a few percent. We find that these models can be 
statistically distinguished from $\Lambda$CDM  through their imprint on the 21-cm angular power spectrum. At the fiducial $z= 1.5$, corresponding to a radio interferometric observation of the post-reionization HI 21 cm observation at frequency $568 \rm MHz$,  these models can infact be distinguished from the $\Lambda$CDM model at $ {\rm SNR }> 3 \sigma$ level using a 10,000 hr radio observation distributed over 40 pointings of a SKA1-mid like radio-telescope. We also show that tracker models are more likely to be ruled out in comparison with $\Lambda$CDM than the thawer models. 
Future radio observations can be instrumental in obtaining tighter constraints on the parameter space of dark energy models and supplement the bounds obtained from background studies.
\end{abstract}

\section{Introduction}
The latest observational data give compelling evidence about the
presence of an unknown dark component with negative pressure in the
universe \citep{supernova,plank13,Sanchez:2012sg}. The contribution of
this unknown component, commonly termed as {\it dark energy}
\citep{2000IJMPD...9..373S,2003RvMP...75..559P,2003PhR...380..235P,2006IJMPD..15.1753C,2002CQGra..19.3435S},
is around $70 \%$ of the total energy budget of the universe. The
presence of such a large unknown component in the universe whose
origin and nature is still unexplained, is a major embarrassment for
cosmologist. Understandably all the future cosmological observations
have a common goal: to know the nature of dark energy.

Cosmological constant (with an equation of state $w=-1$) as proposed
by Einstein himself to obtain a static universe, is the simplest
explanation for the mysterious dark energy, given the fact that a flat
$\Lambda$CDM universe agrees exceptionally well to all the
observational data till date (See also \citep{2015A&A...574A..59D,2014ApJ...793L..40S,2015arXiv150601354T,2016PhRvD..93b3513D} for some recent
contradiction). However,  the problem of extreme fine tuning for the value
of cosmological constant as well as the cosmic coincidence problem,
have inspired researchers to explore beyond the cosmological constant and
study models where dark energy evolves with cosmological evolution.

The natural alternative to cosmological constant is the quintessence
scenario
\citep{Wetterich:1987fm,1988PhRvD..37.3406R,1998PhRvL..80.1582C,1999PhRvD..59b3509L,1999PhRvD..59l3504S,2008PhRvD..77h3515S}
where a minimally coupled scalar field with canonical kinetic term
rolling over a sufficiently flat potential around present time, can
mimic a time varying cosmological constant. Although, one still needs to
do the required fine tuning, one can at least evade  the cosmic
coincidence problem in such a scenario. Various alternatives of
quintessence models such as k-essence
\citep{2000PhRvD..62b3511C,2000PhRvL..85.4438A,2001PhRvD..63j3510A,2002PhRvD..66f3514C,2004MPLA...19..761C,2004PhRvL..93a1301S,2015arXiv151009010S,2016arXiv160205065L},
tachyons
\citep{2003PhRvD..67f3504B,2006JCAP...03..010S,2004PhRvD..69l3517C,2009PhRvD..79l3501A},
non-minimally coupled scalar fields
\citep{2000PhRvD..61f4007B,2002PhRvD..66d3522T,2001PhRvD..63l4006S,2009JCAP...09..027S},
and chameleon fields
\citep{2004PhRvD..69d4026K,2005PhRvD..71d3504W,2008PhRvD..78d3512D}
have also  been widely studied in recent past. A number of phenomenological
potentials have been considered for quintessence field to achieve
$w\approx -1$ in  such a scenario and in all these models, the field
has to slow-roll around present epoch. This sets the mass of the
scalar field to be order of $10^{-33}$ eV. Such a small mass is always
prone to get correction from  various symmetry breakings thereby spoiling
the slow roll conditions for the field. Given the known hierarchy
problem in the standard model, it is extremely difficult to prevent
the small mass of the scalar field to get correction upto the
supersymmetry breaking (SUSY) scale.

This problem has been addressed in the context of string theory by
Panda et al. \citep{2011PhRvD..83h3506P} ( from now on we refer this
model as {\it PST} model) using the idea of axion monodromy in Type-II
B string theory. The resulting potential is a simple linear potential
and the construction is such that the potential does not get
correction upto SUSY breaking scale as the field does not couple to
any standard model sector field ( for details about PST model, please
see \citet{2011PhRvD..83h3506P}). However, there is no mechanism to
prevent the quintessence field to couple with dark matter (DM) sector
which according to the present understanding, has origin in beyond
standard model (BSM) physics.

Coupled quintessence models
where the scalar field is only coupled to DM have been studied in
different contexts starting from background evolution to linear and
nonlinear structure formation \citep{2000PhRvD..62d3511A,2005PhRvD..72d3516K,2006PhRvD..73h3516L,2010PhRvD..82b3528S,2014PhRvD..90h3508A}. Due to the absence of coupling to the
baryonic sector, one can avoid the stringent constraints from local
physics. In most coupled quintessence scenario, the potentials for
scalar field were phenomenological and tracker type. For PST model we
have a thawing scalar field (similar to inflaton) with a linear
potential that is not phenomenological but arises out of the
construction of the model itself. The avoidance of coupling with the
baryonic sector also happens naturally in this set up. The only
phenomenological aspect in this scenario is the form of the coupling
due to our lack of complete understanding about the origin of dark
matter. In recent past, coupled quintessence model in the PST scenario
has been confronted with the latest observational data by Kumar et al
\citep{2013CQGra..30o5011K}. It has been shown that with the current
precision of various observations, a large class of coupled
quintessence behaviour is still indistinguishable from the concordance
$\Lambda$CDM model.

The three dimensional tomographic mapping of the neutral hydrogen (HI)
distribution is a powerful probe to understand large scale structure
formation in the post reionization era \citep{poreion2, poreion0}. The
epoch of reionization was completed by redshift $z \sim 6$
\citep{becker01, fan02}. After this, most of the remnant neutral gas
is contained in the self shielded Damped Ly-$\alpha$ (DLA) systems
\citep{wolfe05}. These are supposedly the primary cosmological source
of HI 21-cm signal \citep{hirev1}. The detection of the individual DLA
clouds is technically very challenging due to their small size and
weakness of the signal ( $< 10 \mu \rm Jy$). But the collective
diffuse HI 21-cm radiation from all the clouds without resolving the
individual DLAs is expected to form a background in radio observations
at frequencies $< 1420 \rm MHz$. Intensity mapping of this background
radiation can yield enormous cosmological information regarding the
background evolution of the Universe as well as the structure
formation in the post-reionization epoch \citep{poreion3, poreion6,
  param1, param2, param3, param4, wyithe08, cosmo14, TGS15}.  The
upcoming Square Kilometer Array (SKA) in various phases has a dominant
science goal of mapping out the large scale distribution of neutral
hydrogen over a wide range of redshifts. Imaging of the Universe using
the redshifted 21-cm signal from redshifts $z \leq 6$ \citep{poreion3,
  poreion2, poreion0} will open new avenues towards our understanding
of cosmology \citep{poreion4, poreion5, poreion6, param1, param2,
  param3}.  The large scale clustering of the HI in the
post-reionization epoch shall directly probe the nature of dark energy
through the imprints of a given model on the background evolution and
growth of structures. As a direct probe of cosmological structure
formation, 21-cm intensity mapping may allow us to distinguish between
dark energy models which are otherwise degenerate at the level of
their prediction of background evolution.

In this paper, we study the prospects of probing coupled quintessence
models using the HI intensity mapping in the context of forthcoming
SKA observations. In addition to the PST model described above, we
also consider other phenomenological potentials that have been
considered in the literature. This gives a detail analysis on future
constraints on coupled quintessence models in the context of
observations from HI 21-cm intensity mapping. We concentrate on the
thawing class of scalar field models ( recent discussions argued that
these are more favored than the tracking ones
\citep{2015PhRvD..91f3006L} ). But to make the investigation complete,
we also consider one particular parametrization (called {\it GCG
  parametrization}) \citep{2012MNRAS.427..988T} that broadly
described both the thawing and tracker models and study the prospects
of distinguishing these two behaviours using the HI intensity mapping
survey by SKA. We also study the clustering on superhorizon scales
where we can no longer ignore the scalar field fluctuations.
\section{Background evolution with scalar field coupled with DM}
We start with a general interacting picture where the DE scalar field is coupledto  the DM sector of the Universe. The visible matter sector (baryons) is not coupled with the DE scalar field. There has been numerous studies in the literature on such ``{\it Coupled Quintessence}" \citep{2000PhRvD..62d3511A,2004PhRvD..69j3524A} models. Here we also follow the same formalism. The relevant equations are given below:
\begin{eqnarray}
 \ddot{\phi}+\frac{dV}{d\phi}+3H\dot{\phi}=C(\phi)\rho_{d} \nonumber \\ 
 \dot{\rho}_{d}+3H(\rho_{d})=-C(\phi)\rho_{d}\dot{\phi} \\
 \dot{\rho_b}+3H(\rho_b)=0 \nonumber\\
  H^2=\frac{\kappa^2}{3}(\rho_b+\rho_{d}+\rho_\phi). \nonumber \\
\end{eqnarray}
\noindent
This is complemented by the flatness condition
\begin{equation}
 1 = \frac{\kappa^2\rho_{b}}{3H^2}+\frac{\kappa^2\rho_{d}}{3H^2}+\frac{\kappa^2\dot{\phi}^2}{6H^2}+
 \frac{\kappa^2V(\phi)}{3H^2}
\end{equation}
Here $C(\phi)$ represents coupling parameter between the scalar field and dark matter. Subscript ``d" represents the DM sector and subscript ``b" represents the baryonic sector. The details of the  physics for the interaction between the dark energy and the dark matter is largely unknown. In view of this, we assume phenomenologically  $C(\phi)$ to be a constant in our subsequent calculations. This is similar to the earlier work by Amendola \citep{2000PhRvD..62d3511A,2004PhRvD..69j3524A} and collaborators on coupled quintessence. For $C=0$ we recover the uncoupled case,hence the system allows us to study both coupled and uncoupled cases.

Next, we construct the following dimensionless variables:
\begin{align}
 x	= \frac{\kappa\dot{\phi}}{\sqrt{6}H}, \hspace{1mm}
 y	= \frac{\kappa\sqrt{V(\phi)}}{\sqrt{3}H} \nonumber \\
 s	=\frac{\kappa\sqrt{\rho_b}}{\sqrt{3}H}, \hspace{1mm}
 \lambda	=	\frac{-1}{\kappa{V}}\frac{dV}{d\phi} \hspace{1mm}
 \Gamma	=	\frac{V\frac{d^2V}{d\phi^2}}{\left(\frac{dV}{d\phi}\right)^2} 
\end{align}

\noindent
Note that the parameter $\Gamma$ is related to the form of the potentials in our model. For the PST model described in the introduction, the potential is linear, hence $\Gamma$ vanishes. For completeness, we also consider other power-law potentials of the form $V(\phi) \propto \phi^n$ where $\Gamma = (n-1)/n$.

The density parameter $\Omega_{\phi}$ and the equation of state for the scalar field $w_{\phi}$ can be written in terms of $x$ and $y$ as:
\begin{align}
 \Omega_\phi	=&	x^2+y^2 \\
 \gamma	=&	1+w_\phi	=	\frac{2x^2}{x^2+y^2}
\end{align}

\noindent
With this, one can form an autonomous system of equations:
\begin{align}
 \Omega_\phi'	=&	W\sqrt{3\gamma\Omega_\phi}(1-\Omega_\phi-s^2)+3\Omega_\phi(1-\Omega_\phi)(1-\gamma) \nonumber \\
 \gamma'	=&	W\sqrt{\frac{3\gamma}{\Omega_\phi}}(1-\Omega_\phi-s^2)(2-\gamma)+\lambda\sqrt{3\gamma\Omega_\phi}(2-\gamma)\nonumber\\
 - &3\gamma(2-\gamma) \nonumber \\
 s'	=&	-\frac{3}{2}s\Omega_\phi(1-\gamma) \nonumber \\
 \lambda'	=&	\sqrt{3\gamma\Omega_\phi}\lambda^2(1-\Gamma),
\end{align}
where $W=\frac{C}{\kappa}$. We evolve the above system of equations
from the decoupling era ($a=10^{-3}$) to the present day ($a=1$). We
need to fix the initial conditions for $\gamma$, $\Omega_{\phi}$ $z$
and $\lambda$ to solve the system of equations. For thawing models, scalar field is
initially frozen due to large Hubble damping, and this fixes the
initial condition $\gamma_{i} \approx 0$.  The initial value
$\lambda_{i}$ is a model parameter; for smaller $\lambda_{i}$, the
equation of state $w_{\phi}$ for the scalar field always remain close
to cosmological constant $w=-1$ whereas for larger values of
$\lambda_{i}$, $w_{\phi}$ increases from $-1$ as the universe
evolves. The contribution of scalar field to the total energy density
is negligibly small in the early universe (except for early dark
energy models that we are not considering in this study). But we need
to fine tune it initially in order to obtain a correct value of
$\Omega_{\phi}$ at present. This fixes the initial condition for
$\Omega_{\phi i}$. Similarly, we need to fix the initial value of $s$ (
which is related to the density parameter for baryons) to get right
value of the $\Omega_{b}$ at the present epoch. In our subsequent calculations we fix
$\Omega_{b0} = 0.05$.
\section{Growth of Matter Fluctuations in the Linear Regime}
We next study the growth of matter fluctuations in the linear
regime. Here matter consists of both dark matter and baryons; but in
the late universe ( which is the time period we are interested in),
the dark matter perturbation is dominant and baryons follow the dark
matter perturbation. Hence we ignore the baryonic contribution in our
calculations. We should stress that even if we include the baryon
contribution (which is very straightforward to do), our results do not
change.  We work in the longitudinal gauge:
\begin{equation}
ds^{2} = a^{2}\left[-(1+2\Phi)d\tau^2 + (1-2\Psi) dx^{i}dx_{i}\right],
\end{equation}
\noindent
where $\tau$ is the conformal time and $\Phi$ and $\Psi$ are the two gravitational potentials. In the absence of any anisotropic stress $\Phi= \Psi$.
We follow the prescription by Amendola \citep{2000PhRvD..62d3511A,2004PhRvD..69j3524A} and write the equations for the perturbations in dark matter density in the Newtonian limit. This is valid assumption for sub horizon scales. In these scales, one can safely ignore the clustering in the scalar field. Under these assumptions, the linearized equations governing the growth of fluctuations in dark matter is given by:
\begin{equation}
 \delta''_{d}+\left(1+\frac{\mathcal{H}'}{\mathcal{H}}-2\beta_{d}x\right)\delta'_{d}-\frac{3}{2}(\gamma_{dd}\delta_{d}
 \Omega_{d})=0.
\end{equation}
\noindent
The prime denotes differentiation w.r.t to $\log a$. $x$ is given by equation (4). Here $\beta_d=W$, $\gamma_{dd}=1+2\beta_{d}^2$, ${\mathcal{H}}$ is the conformal Hubble parameter ${\mathcal{H}=aH}$ and $\delta_{d}$ is the linear density contrast for the DM. We solve the equation with the initial conditions $\delta_{d} \sim a$ and $\frac{d\delta_{d}}{da} = 1$ at decoupling $a \sim 10^{-3}$. This is valid since the universe is matter-dominated at the epoch of decoupling. 
We take the Fourier transform of the above equation and define the linear growth function $D_{d}$ and the linear growth rate $f_{d}$ as
\begin{eqnarray}
\delta_{d k} (a) &\equiv& D_{d}(a)\delta_{d k}^{ini}\\
f_{d} &=& \frac{d \ln{D_{d}}}{d \ln{a}}.
\end{eqnarray}
The linear dark matter power spectrum defined as
\begin{equation}
P(k,z) = A_0 k^{n_s} T^2(k) D_{d n}^2(z).
\end{equation}
Here $A_0$ is the normalization constant fixed by $\sigma_{8}$ normalization, $n_{s}$ is spectral index for the primordial density fluctuations generated through inflation, $D_{d n} (z)$ is growth function normalized such as it is equal to unity at $z=0$ i.e. $D_{d n}(z)=\frac{D_{d}(z)}{D_{d}(0)}$ and $T(k)$ is the transfer function as prescribed by Eisenstein and Hu \citep{EHU}.
\section{ The redshifted 21 cm signal from the post-reionization epoch}
The neutral hydrogen (HI) distribution in post-reionization epoch is modeled by a mean neutral fraction ${\bar x}_{\rm HI}$ which remains constant over a wide redshift range $ z \leq 6$ \citep{xhibar1, xhibar2} and a linear bias parameter $b_T$ which relates the HI fluctuations to the fluctuations in  the underlying dark matter distributions \citep{bagla2, tgs2011}. The quantity of interest is the fluctuation of the excess HI 21 cm brightness temperature $\delta T_b$. Denoting the comoving distance to the redshift $z$ by $r$ we have $\delta T_b$ given by a fluctuation field on the sky corresponding to  radial and angular coordinates $( z,  r {\hat {\bf n}}) $ as $\delta T_b = \bar {T}(z) \times \eta_{\rm HI} (r {\hat {\bf n}})$ \citep{bharad04} where,  
\begin{equation}
\eta_{HI}( r {\bf\hat{n}}, z)=\bar{x}_{HI}(z)\left[\delta_{HI}(z,{\bf\hat{n}})-\frac{1+z}{H(z)}\frac{\partial v}{\partial r}\right].
\end{equation} and
\begin{equation}
\bar T(z)=4.0 {\rm mK} (1+z)^{2}\left(\frac{\Omega_{b0}h^{2}}{0.02}\right)\left(\frac{0.7}{h}\right) \left( \frac{H_{0}}{H(z)}\right)
\end{equation}
Here, $\delta_{H}$ denotes the HI fluctuations and $v$ denotes the peculiar velocity of the gas. If $\Delta({\bf k}, z)$ and $\Delta_H({\bf k}, z) $ denote dark matter overdensity $\delta_d$  and $\delta_{H}$ respectively in  Fourier space then they are related by a bias function $b_T({\bf k}, z)$ as  $\Delta_H({\bf k}, z) =b_T({\bf k}, z) \Delta({\bf k}, z)  $. On large scales of our interest, the bias is found to be a constant in numerical simulations of the post-reionization HI signal \citep{tgs2011}. We use a linear bias model in this analysis.
If the peculiar velocities of the gas are sourced by dark matter over densities then the angular power spectrum of the brightness temperature in the flat sky limit is given by \citep{datta1} 
\begin{equation} 
C_{l} = \frac{{\bar T}^2 \bar x_{HI}^2 b_T^2}{\pi r^2}\int
\limits_0^\infty dk_{\parallel} ~( 1 + \beta \mu^2)^2~ P(k,
z) \end{equation} where $P(k, z)$ is the dark matter power spectrum defined in equation (12), $
k = \sqrt{k_{\parallel}^2 + \frac{l^2}{r^2}}$, $\mu =
\frac{k_{\parallel}}{k}$ and $\beta = \frac{f_{d}(z)}{b_T}$ with $f_{d}(z)$
is the growth factor for dark matter defined in equation (11).  We note that the
redshift dependent quantities $\bar T$, and $r$ are directly related
to the background cosmology and $\beta$ and $P(k, z)$ imprint both
background history and structure formation. We adopt the value of $\Omega_{HI} = 10^{-3}$ at $ z
<3.5$. This yields $\bar x_{HI} = 2.45\times 10^{-2}$ \citep{xhibar1} which is assumed
to be constant across the redshift range of our interest.

We consider a radio interferometric measurement of the power spectrum of the
21 cm brightness temperature.  The directly measured  'Visibility' is a 
function of frequency $\nu$ and baseline ${\bf U} = {\bf k_{\perp}} r/ 2 \pi$ and  allows us to compute the angular power spectrum $C_l$ directly using Visibility-Visibility correlation \citep{bali} with the association $ l = 2\pi U$.
 
The noise in the measurement of angular power spectrum comes from cosmic variance on small scales and instrument noise on smalls scales.
We have
\begin{equation}
{\Delta C_l} = \sqrt{\frac{2}{(2l + 1)\Delta l f_{sky} N_p}} \left ( C_l + N_l \right )
\end{equation}
where $N_p$ denotes the number of pointings of the radio interferometer, $f_{sky}$ is the fraction of sky observed in a single pointing, and $\Delta l $ is the width of the $l$ bin.
The noise power spectrum $N_l$ is given by 
\begin{equation}
N_l = \left( \frac{\lambda^2}{A_d} \right)^3\frac{T_{syst}^2}{N_{pol} \Delta \nu t_o n({\bf U})}
\end{equation}

where, $\lambda$ is the observed wavelength, $A_d$ is the effective antenna diameter, $T_{syst} $ is the  system temperature, $t_0$ is the observation time, $N_{pol} $ denote the number of polarization states used, $\Delta \nu $is the frequency band and $n({\bf U}) $ is the baseline distribution function normalized as
\begin{equation}
\int d^2 {\bf U}~~ n({\bf U}) = \frac{N_{ant}( N_{ant} -1) }{2}
\end{equation}
where $N_{ant}$ denotes the number of antennae in the radio array.

We consider a radio interferometer with parameters roughly following
the specifications of SKA1-mid \footnote{http://www.skatelescope.org/wp-content/uploads/2012/07/SKA-TEL-SKO-DD-001-1\_BaselineDesign1.pdf}.  The fiducial redshift $z= 1.5$
corresponds to an observing frequency of $568{\rm MHz}$ which falls in the
band of frequencies to be probed by SKA1-mid. We consider a frequency
bandwidth of $32 {\rm MHz}$ around the central frequency. The array is
assumed to be composed of 200 dishes each of diameter $15$m. The
antennae are distributed in a manner such that $75\%$ of the dishes
are within $2.5$Km radius and the density of antennae are assumed to
fall off radially as $ r^{-2}$.
We also use $ T_{syst} = 180\left (\frac{\nu}{180{\rm MHz}} \right) ^{-2.6}$ K. in our error estimates. With this we calculate the error bar $\Delta C_{l}$ for SKA1-mid assuming a fiducial $\Lambda$CDM model with $\Omega_{\Lambda} = 0.7$, $\Omega_{b0} = 0.05$, $n_{s} = 1$, $h=0.7$ and $\sigma_{8} = 0.8$. We also fix the constant linear bias to be $1.0 $ at the fiducial redshift from numerical simulations of the 21cm signal \citep{tgs2011}.

\section{Results}
\begin{figure*}
\begin{center} 
\resizebox{200pt}{160pt}{\includegraphics{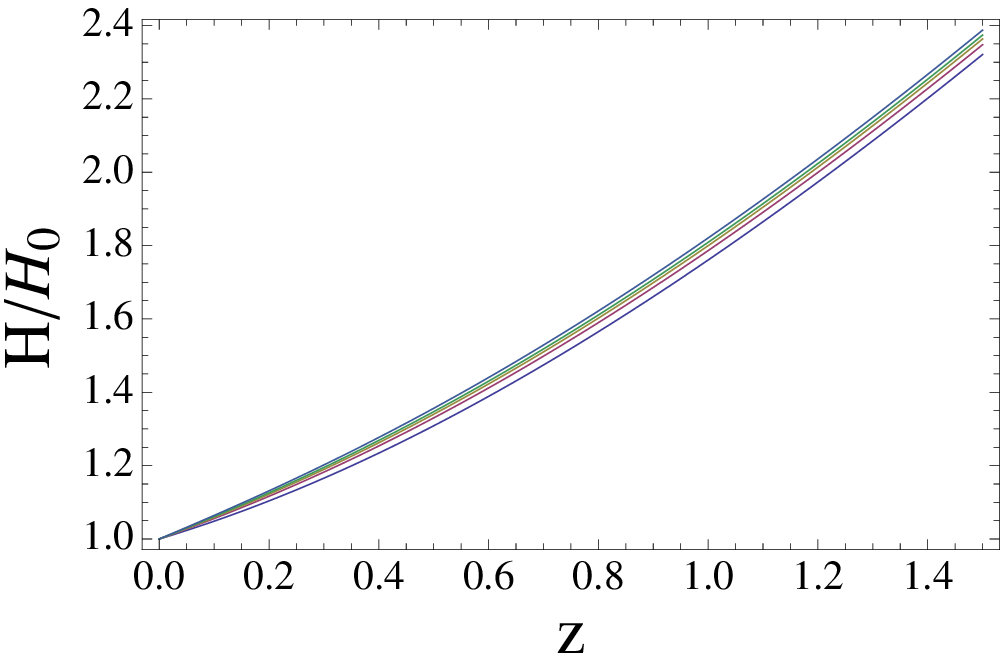}}
\hspace{1mm} \resizebox{200pt}{160pt}{\includegraphics{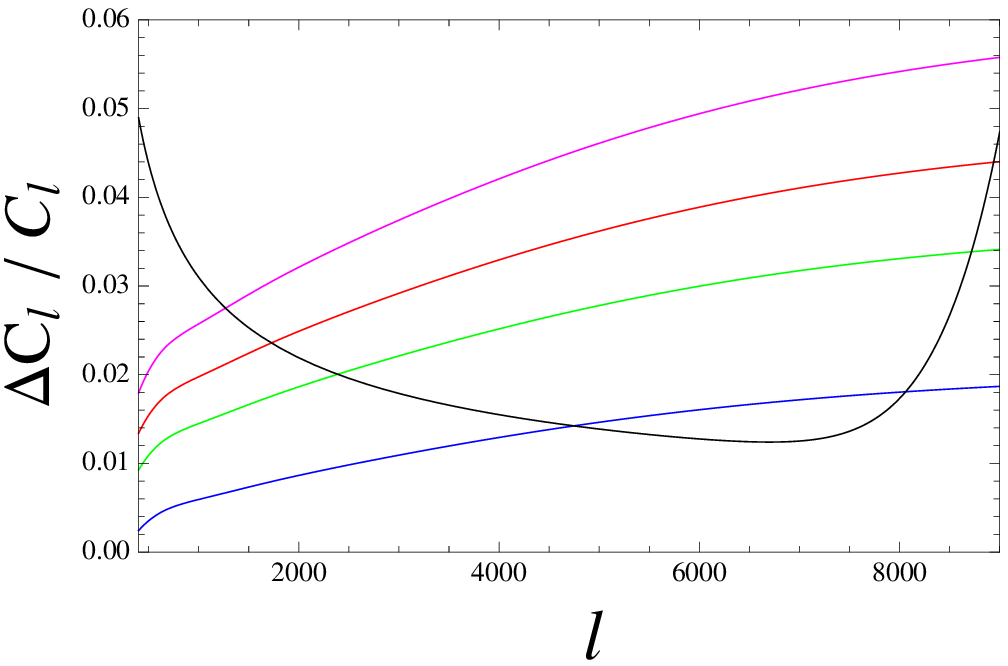}}
\end{center}
\caption{({\it left}) The evolution of normalized Hubble parameter
  with redshift for $\Lambda$CDM and different combinations ($W=0.04,
  \lambda_{i} = 0.5$), ($W=0.05, \lambda_{i} = 0.5$), ($W=0.06,
  \lambda_{i} = 0.5$), ($W=0.07, \lambda_{i} = 0.5$), ($W=0.08,
  \lambda_{i} = 0.5$) from bottom to top for linear potential $V \sim
  \phi$. $\Omega_{m0} = 0.3$ for all the plots. ({\it right})  From bottom to top, the relative difference of the 21-cm angular power spectrum from the fiducial $\Lambda$CDM at a redshift $ z=1.5$. The same combinations of ($W, \lambda_{i}$) as in left figure are used. The black line corresponds to the noise level with SKA1-mid like telescope assuming the fiducial  $\Lambda$CDM value model.  }
\end{figure*}

In figure 1 (left), we plot the evolution of Hubble parameter $H(z)$
with redshift for different combinations of $W$ and $\lambda_{i}$ and
also for $\Lambda$CDM for the PST model with linear potential.
Remember $W$ parameter determines the strength of the coupling and
$\lambda_{i}$ determines the deviation from the cosmological
constant. We choose the combinations ( mentioned in figure 1) so that
the deviation from $\Lambda$CDM is very small and in actual it is
around $3-4\%$ or less upto a redshift $z=1.5$. This is much smaller than
the current error bar in the measurement of $H(z)$ at different
redshift.

The fact that the models are statistically indistinguishable at the
level of their predictions about background evolution, prompts us to
investigate their signature in structure formation. The imprint of the
dark energy model on the 21-cm angular power spectrum is not only
through the background model contained in $H(z)$ and $r$, but also
through the manner in which they affect the growth of structures and
thereby affecting the matter power spectrum.  We consider a $10,000$
hrs radio observation of the 21-cm signal using a SKA1-mid like radio
telescope where the total observation time $t_0$ is distributed over
40 radio pointings of individual $250 hrs$ observation.  The power
spectrum is binned over $l$ with $\Delta l = l/5$. The noise level
shows a steep rise at low multipoles owing to cosmic variance and also
at large multipoles due to dominant instrument noise.  In figure 1
(right), we show the deviation of the dark energy models models from
$\Lambda$CDM universe in terms of the angular power spectra for HI
intensity mapping. It is clearly visible that most of these models can
be ruled out in comparison with $\Lambda$CDM model with future
SKA1-mid data in an intermediate multipole region around $l \sim 7000$
at a  $3\sigma$ to $5\sigma$ confidence level.

\begin{figure*}
\begin{center} 
\resizebox{200pt}{160pt}{\includegraphics{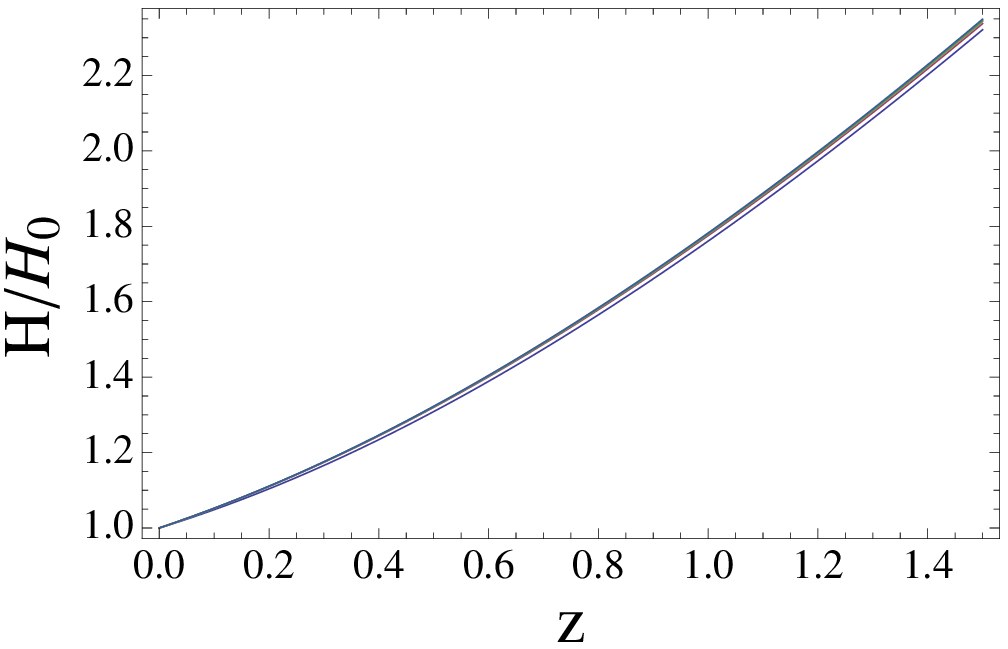}}
\hspace{1mm} \resizebox{200pt}{160pt}{\includegraphics{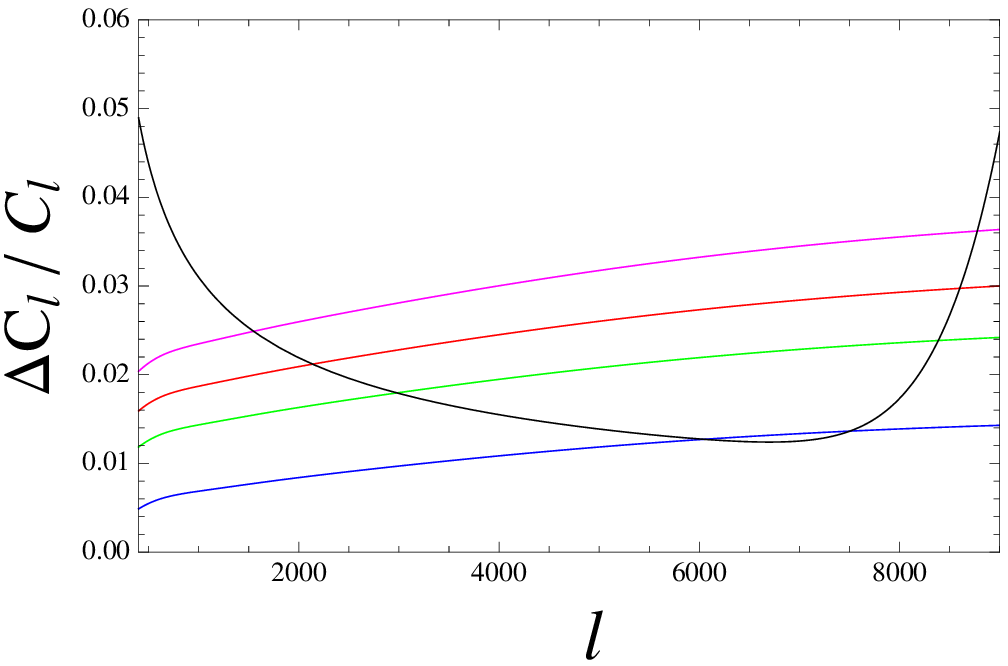}}
\end{center}
\caption{Same as in figure 1 but for inverse power law potential $V \sim \phi^{-2}$.} 
\end{figure*}

\begin{figure*}
\begin{center} 
\resizebox{200pt}{160pt}{\includegraphics{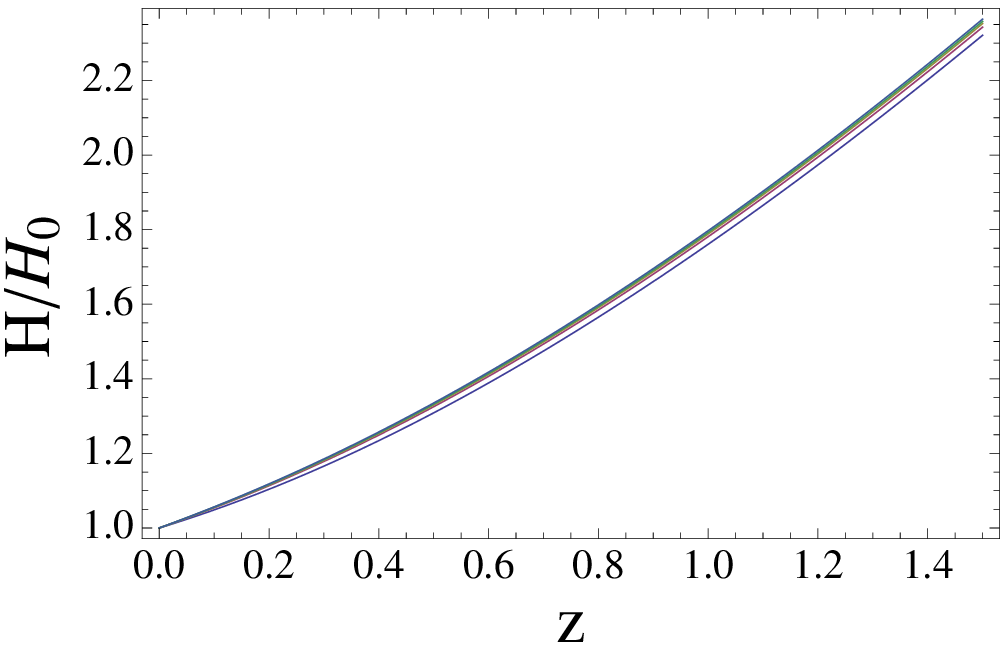}}
\hspace{1mm} \resizebox{200pt}{160pt}{\includegraphics{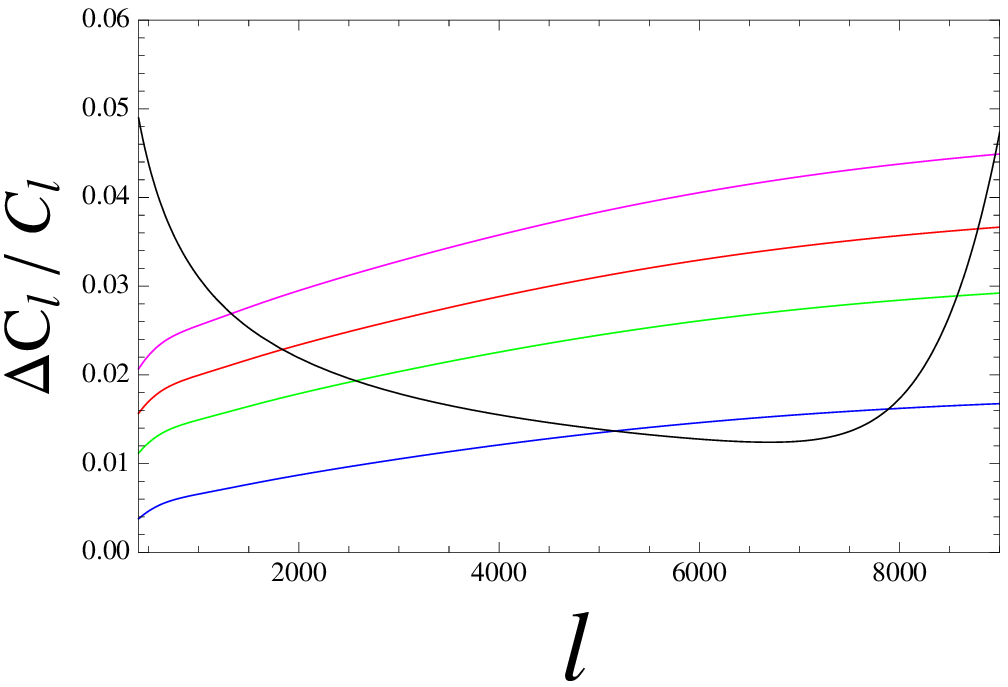}}
\end{center}
\caption{Same as in figure 1 but for for square law potential $V \sim \phi^{2}$.} 
\end{figure*}

In figure 2 and 3, we plot the same but for scalar field potentials $V
\sim \phi^{-2}$ and $V \sim \phi^{2}$. One can see, in these cases the
confidence level at which one can distinguish these models with
$\Lambda$CDM, decrease slightly but still one can distinguish them
from $\Lambda$CDM at $3\sigma$ confidence level or more. Note that we
choose combinations of parameters $W$ and $\lambda_{i}$ for which the
Hubble parameter deviates from $\Lambda$CDM value by $3-4\%$ which is
very conservative choice. Current data allows bigger deviations from
$\Lambda$CDM value. In  such cases, the deviation in $C_{l}$ from
$\Lambda$CDM predictions should be certainly distinguished with larger confidence
level with future SKA1-mid data.

Also if we put $W=0$ in all these cases, the fractional difference with $\Lambda$CDM will be much less than the error bar and can not be distinguished at all from $\Lambda$CDM with future survey like SKA1-mid.

\section{Thawing Vs Tracker}

In the previous sections, we consider scalar field models which are {\it thawer} in nature. In such models, the scalar field is initially frozen due to large hubble damping and the equation of state of the scalar field is very close to $-1$. As the universe evolves, hubble damping decreases and the scalar field slowly thaws away from the frozen state and the equation of state of the scalar field slowly increase towards $w > -1$. There is another class of models, known as the {\it tracker models} where initially the scalar field fast rolls due to the steep nature of the potential and mimics the background matter density ($w \sim 0$). In late times, the scalar field potential flattens up and the scalar field finally freezes to $w \sim -1$ behaviour. Although a variety of potentials can give rise to both thawer and tracker potentials, it may be useful to have simple parametrization for the equation of state of the scalar field that broadly describes these two behaviours. The generalized chaplygin gas (GCG) equation of state described by $p = -A/\rho^{\alpha}$ where $A$ and $\alpha$ are two constant parameters, is useful for this purpose \citep{2012MNRAS.427..988T}. For such a parametrization the dark energy equation of state is given by

\begin{equation}
w_{de} = -\frac{A_{s}}{A_{s} + (1-A_{s}) a^{-3 (1+\alpha)}},
\end{equation}

\noindent
where $A_{s} = \rho_{de 0}/A^{1+\alpha}$. It is straightforward to check that for $(1+\alpha) < 0$, $w_{de}$ behaves like thawer model while for $(1+\alpha) > 0$, $w_{de}$ behaves like tracker model. The parameter $A_{s}$ is related to the current value of the equation of state for the dark energy, $A_{s} = -w_{de 0}$. In figure 4, we show these two behaviours.

\begin{figure*}
\begin{center} 
\resizebox{200pt}{160pt}{\includegraphics{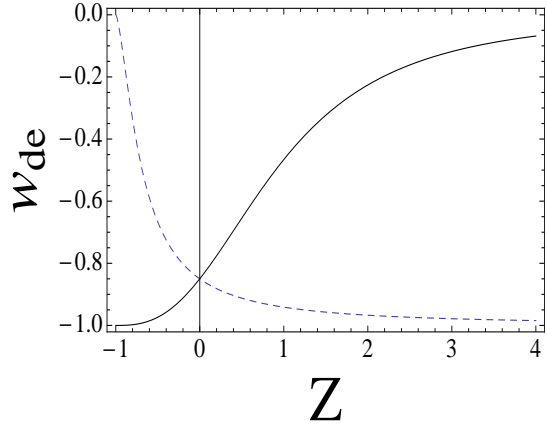}}
\end{center}
\caption{Thawing and Tracker behaviour for the dark energy equation of state. $\alpha = -1.5$ for dashed and $\alpha = -0.1$ for solid line. $A_{s} = 0.85$ for both the lines. } 
\end{figure*}

With this, we investigate whether one can distinguish these two behaviours from $\Lambda$CDM using the future SKA data. We concentrate on the uncoupled case where the dark energy is not coupled with the dark matter and follow the same procedure as described in section 3 and 5. The result is shown in figure 5. 

\begin{figure*}
\begin{center} 
\resizebox{200pt}{160pt}{\includegraphics{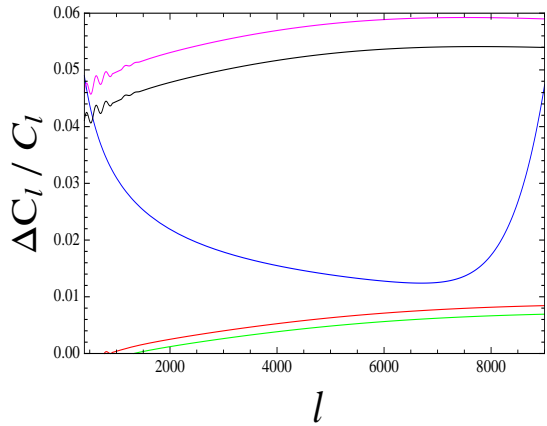}}
\end{center}
\caption{The blue line is for error bar for SKA-mid. The lower two lines for thawer model with $\alpha = (-1.1, -1.05)$ respectively from top to bottom and upper two lines are for tracker models with $\alpha = (-0.05, -0.1)$ from top to bottom. $A_{s} = 0.95$ for all the lines. } 
\end{figure*}

The result for the thawer model is consistent with our previous observation in section 5 that without interaction with DM  ($W=0$), the thawing model can not be distinguished from $\Lambda$CDM with future SKA-mid. On the other hand, the tracker model can be distinguished from $\Lambda$CDM with very high confidence with future SKA observations even without the interaction with DM. 

\section{Fluctuations in the Super-Horizon scales} 

In the previous sections, we study the dark matter density perturbation for quintessence models in the sub-horizon scales where we ignored the perturbations in the dark energy scalar field and concentrate in the Newtonian limit. We show that it will be possible to distinguish these scalar field models from $\Lambda$CDM specially for coupled scenario using the observations of angular power spectra from HI 21-cm mapping survey like SKA-mid. In this section, we study how these models deviate from $\Lambda$CDM at super-horizon scales. On such scales, we can not ignore the perturbations in the scalar field for dark energy and also one has to do the full relativistic perturbation. Such study has been done by various authors in recent past. Here we specifically concentrate on the coupled quintessence of thawer class with linear potentials as constructed in the PST model. It can be straightforwardly generalized for any other scalar field potentials.

For this purpose we follow the set up as provided by Unnikrishnan et al \citep{2008PhRvD..78l3504U}. We work in the longitudinal gauge as described by the metric in equation (8) with $\Phi = \Psi$ for vanishing anisotropic stress. The perturbed Einstein equations are given $\delta G_{\mu\nu} = \delta T_{\mu\nu}$, where $\delta T_{\mu\nu}$ contains both the pertubations in the matter part as well as in the scalar field part. These are complimented by the equations arising from the Bianchi identity $T^{\mu}_{\nu ; \mu} = 0$ where $T^{\mu}_{\nu}$ contains the unpertubed part as well as the pertubed part from the matter and scalar field. We refer the reader \citep{2008PhRvD..78l3504U} for the detail calculations. We generalize the results in \citep{2008PhRvD..78l3504U} for the coupled scenario. The final relevant equations are given by:

\begin{eqnarray}
&& \Phi_N^{\prime \prime} + \left[\frac{h^{\prime}}{h}-\frac{3}{(1+z)}\right]\Phi_N^{\prime} + \left[ \frac{3}{(1+z)^2} - 2\frac{h^{\prime}}{h}\frac{1}{(1+z)}\right]\Phi_N = \frac{1}{2}\left[\tilde{\phi}^{\prime}\delta\phi_N^{\prime} - \Phi_N\tilde{\phi}^{\prime 2} - \frac{\delta \phi_N}{\tilde{\phi}^{\prime}}\frac{\tilde{V}^{\prime}}{h^2(1+z)^2} \right]  \\
&&\delta\phi_N^{\prime\prime} + \left[\frac{h^{\prime}}{h} - \frac{2}{(1+z)} \right]\delta \phi_N^{\prime} + \frac{k^2}{H_0^2}\frac{\delta \phi_N}{h^2} - 4\Phi_N^{\prime}\tilde{\phi}^{\prime} -2\Phi_N \left[\left(\frac{h^{\prime}}{h} - \frac{2}{(1+z)}\right)\tilde{\phi}^{\prime} + \tilde{\phi}^{\prime \prime} \right]+ \frac{1}{h^2(1+z)^2}\delta\phi_N\frac{d^2\tilde{V}}{d\tilde{\phi}^2}  \\
 &&=-3W\frac{\Omega_m(z)}{(1+z)^2}\frac{\delta_c}{\Phi_i} \\
 &&\frac{\delta_c^{\prime}}{\Phi_i} + \frac{(1+z)}{3\Omega_m(z)}\frac{k^2}{H_0^2}\frac{1}{h^2}\left[2(1+z)\Phi_N^{\prime} - 2\Phi_N^{\prime} - (1+z)\phi^{\prime}\delta\phi_N \right]+3\Phi_N^{\prime} =-W\delta\phi_N^{\prime}	
\end{eqnarray}

\begin{figure*}
\begin{center} 
\resizebox{200pt}{160pt}{\includegraphics{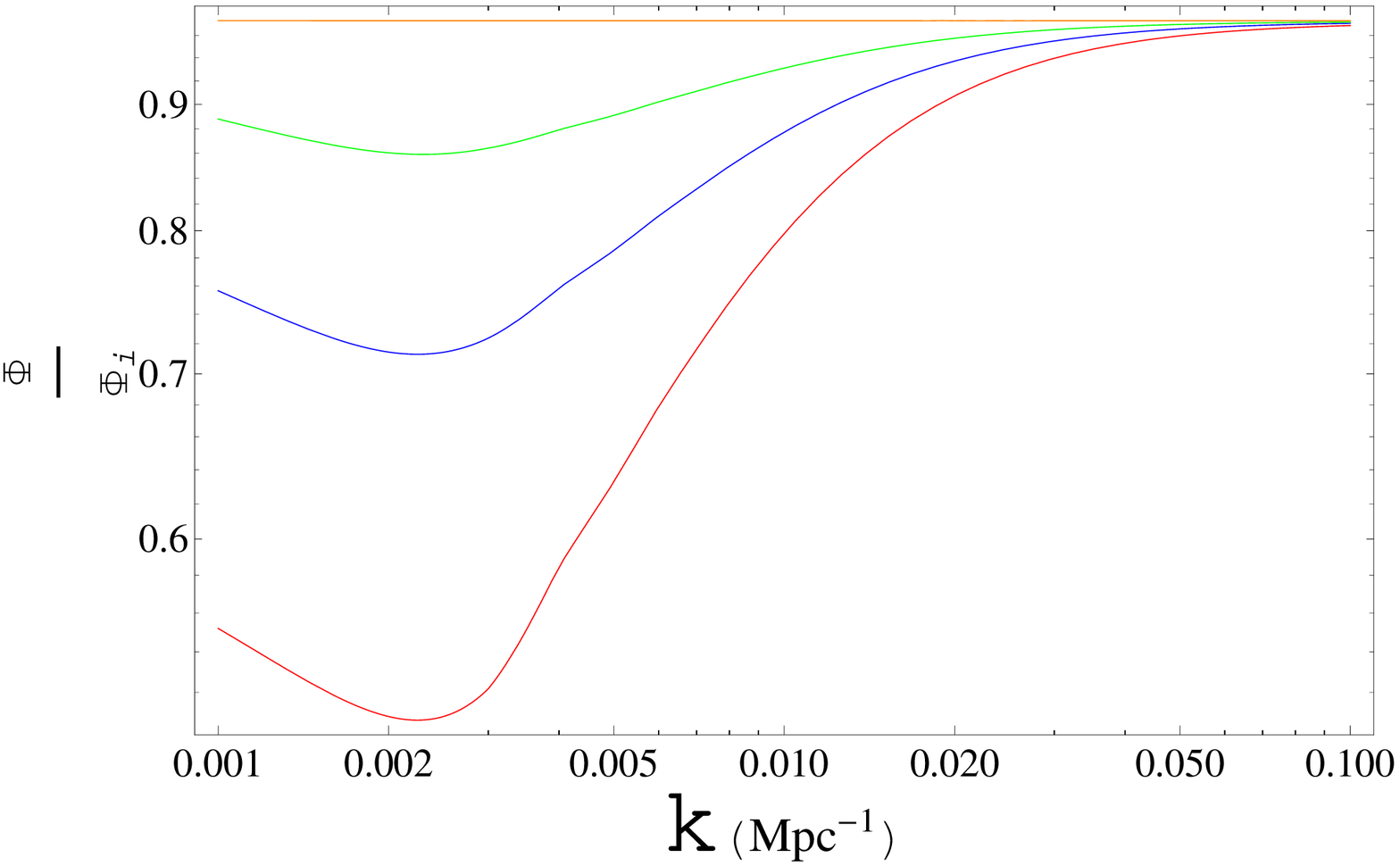}}
\hspace{1mm} \resizebox{200pt}{160pt}{\includegraphics{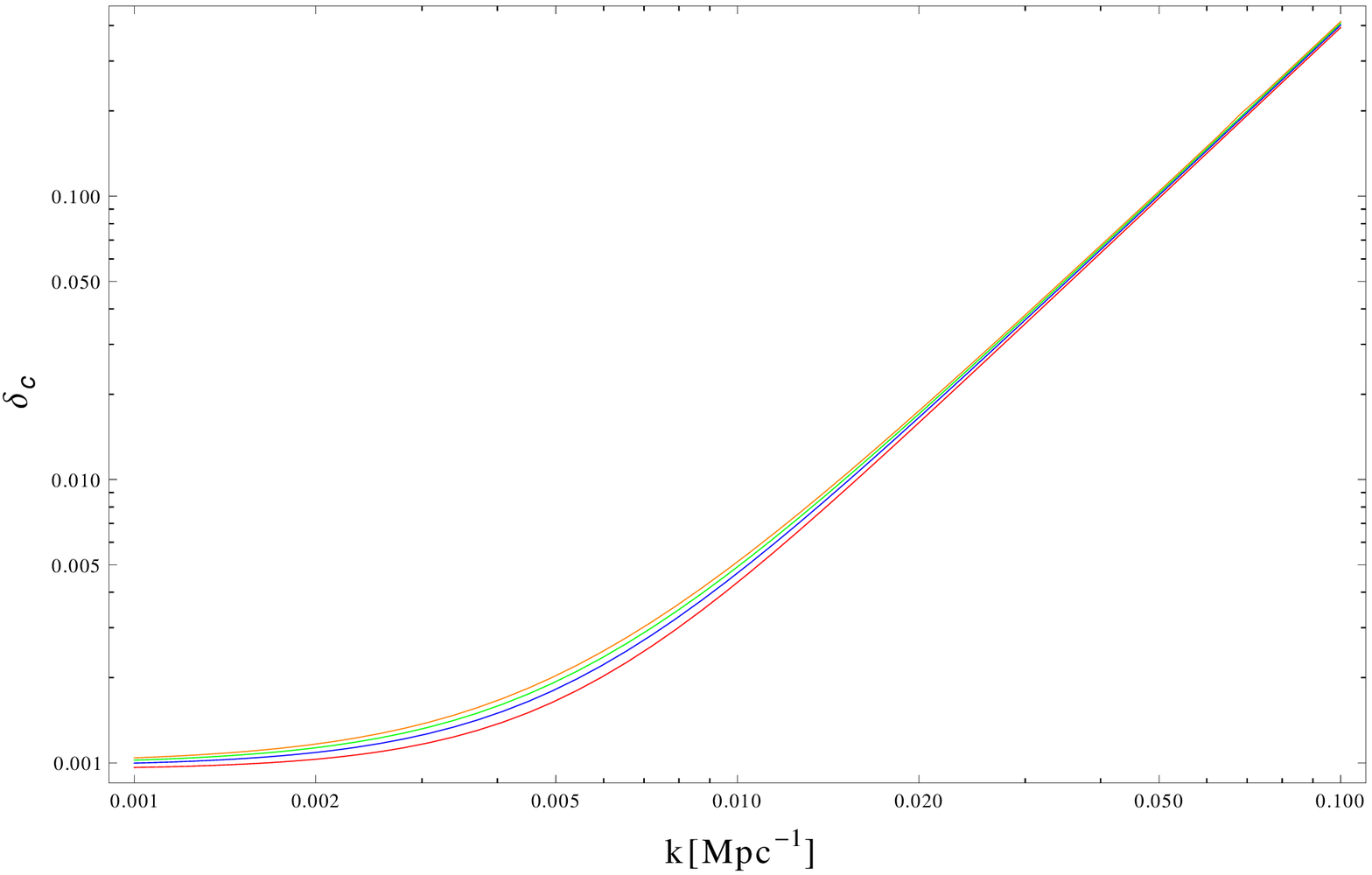}}\\
\resizebox{200pt}{160pt}{\includegraphics{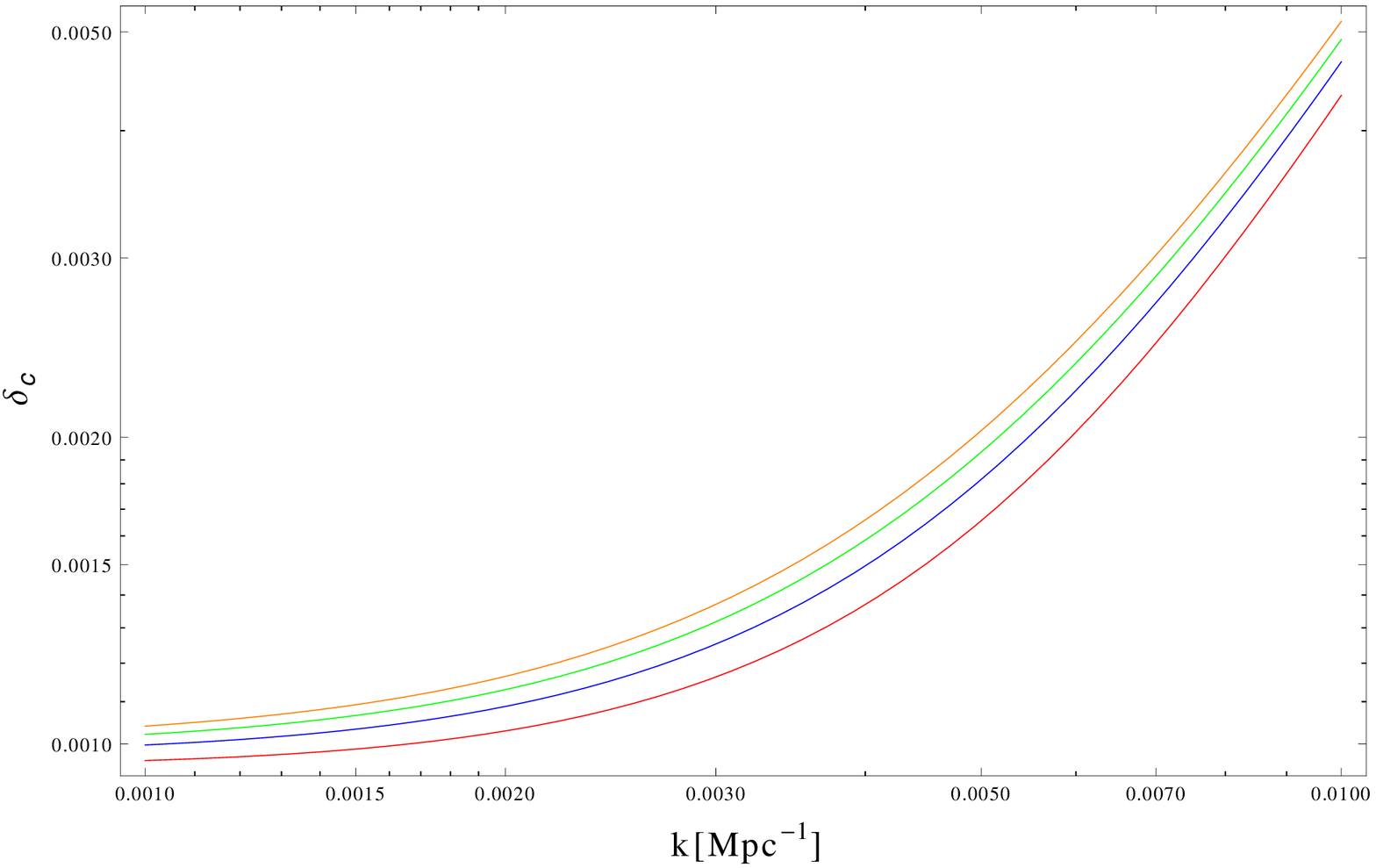}}
\end{center}
\caption{(upperleft): Scale dependence of gravitational potential at $z=1.5$. (upperight): scale dependence of $\delta_{c}$ at $z=1.5$. (bottom): Magnified plot for $\delta_{c}$ between $k= 0.001$/Mpc and $k=0.01$/Mpc. For all the curves: topmost is for $\Lambda$CDM and $W= 0.04,0.06,0.08$ from top to bottom. $\Lambda_{i} = 0.5$ for all the plots.} 
\end{figure*}

\noindent
Here the inhomogeneous scalar field is $\phi({\vec{x}},t) =
\tilde{\phi}(t) + \delta\phi({\vec{x}},t)$ and $\delta
\phi_N=\frac{\sqrt{8 \pi G \delta \phi}}{\Phi _i}$, $\Phi_{N} =
\frac{\Phi}{\Phi_{i}}$, $\Phi_{i}$ being the value of $\Phi$ initially
and $\delta_{c}$ is the density contrast for the dark matter
fluctuations, $h = H/H_{0}$ for the background evolution and ``prime''
is differentiation w.r.t redshift.  Also $\tilde{V}=\frac{8\pi
  G}{H_0^2}V$. The initial conditions at decoupling ($a \sim 10^{-3}$)
are chosen in the following way: $\Phi_{N} = 1$ and $\Phi_{N}^{\prime}
= 0$; $\delta\phi = \delta\phi ^{\prime} = 0$ ( scalar field is
homogeneous initially) and $\delta_{c} \sim a$. In figure 6, we show
the scale dependence of the gravitational potential $\Phi(k)$ and the
DM density contrast $\delta_{c}(k)$. We fix the redshift at $z=
1.5$. It is evident that on large scales, there are large deviations
from $\Lambda$CDM model in the coupled scenario as we take into account
the perturbations of the scalar field. On smaller scales (increasing $k$),
however, the effect of scalar field perturbation becomes negligible. One should note here,  that observing large scale effect of the scalar field
dark energy perturbation is difficult due to large cosmic
variance. It is possible, in principle to
detect the imprint of such large scale effects by considering a full
sky 21-cm survey covering a large band in say $ 0.5 < z < 3$ and
collapsing all the multipoles in an experiment where instrumental noise  is made to go below the cosmic variance level. Large survey volumes may in future
allow us to detect the imprints of clustering dark energy using the
maximally available tomographic 21-cm data.

\section{Discussions}

The large scale clustering of the neutral hydrogen in the post
reionization era contains a lot of information about our universe for
both the background evolution as well as formation of large scale
structures. Hence it is a natural probe for dark energy behaviour. In
this paper we study the prospects of probing a large class of scalar
field dark energy models using angular power spectra for the HI 21-cm
intensity mapping from future SKA like instruments.  Several
observational challenges poses serious difficulties towards the
detection of the cosmological redshifted 21-cm signal.  Astrophysical
foregrounds from galactic and extra galactic sources are several
orders of magnitude larger than the signal \citep{ghosh2011} and
significant amount of foreground subtraction is required for a
statistical detection of the signal \citep{ fg4, fg1, fg5, fg9,
  ghosh2011, alonso2014}.  The cross correlation of the redshifted 21
cm signal with other cosmological probes like the Lyman-alpha forest
and Lyman-Break galaxies, has been proposed \citep{tgs5, TGS15,
  navarro2} to cope with the effect of foreground residuals. Further
man made Radio frequency interferences and other systematic effects
like calibration errors shall also have to be tackled before obtaining
the pristine cosmological signal.

 In this work we have considered thawing class of coupled quintessence
 models with different potentials including that constructed in a
 string theory set up in the PST model. The equations are constructed
 in such a way that one can easily switch off the interaction term and
 study the uncoupled case as well. We show that models which deviate
 from the $\Lambda$CDM universe at $3-4\%$ level and can not be
 distinguished by current observations, can be easily be ruled out in
 comparison with $\Lambda$CDM model in an multipole region $l \sim
 7000$ with $3-5\sigma$ confidence level which is very
 encouraging. But with the anticipated error bar for SKA1-mid, it will
 still not be possible to distinguish the uncoupled models from
 $\Lambda$CDM. Large survey volumes in future however may allow a
 possible detection.

Although our analysis focuses primarily on the thawing class of scalar field
models, tracker class of models is  another possibility. To distinguish
between these classes of models without considering individual
potentials, we consider the GCG equation of state which broadly
describe both the models for different parameter ranges. With this, we
compare the thawing and tracking class of models for the uncoupled
case and show that tracker models can be easily ruled with very high
confidence level in comparison with both $\Lambda$CDM as well as
thawing models. The same is true for coupled tracking model although
we do not show it explicitly.

In the end, we have studied the deviations of the coupled quintessence
model from $\Lambda$CDM on very large scales where one can no longer
ignore the perturbations in the scalar field and one needs to
consider the full relativistic calculations. We show that there is
substantial deviation from $\Lambda$CDM on large scales for coupled
quintessence. But limitation due to cosmic variance on large scales is
a problem to probe these deviation. Future tomographic 21-cm data with large survey volume may be useful to probe dark energy on these scales.
\section{Acknowledgements}
TGS would like to acknowledge the Department of Science and Technology (DST), Government of India for providing financial support through the project SR/FTP/PS-172/2012. AH acknowledges UGC, Govt of India for financial support. ST acknowledges financial support from IISER, Mohali where part of the work has been done. ST also thanks CTP, JMI for providing research facilities. 
\bibliographystyle{mnras}
\bibliography{references}

\end{document}